\documentclass[useAMS,usenatbib]{mn2e}
\usepackage{amsmath}
\usepackage{graphicx}

\usepackage{ enumitem, tabto, xfrac}
\usepackage[]{units}
\usepackage{tabu}
\usepackage{morefloats}
\usepackage{float}
\usepackage{amssymb}
\usepackage{xcolor}
\usepackage[normalem]{ulem}

\title[Spectral Index Measurements with EDGES]{Improved Measurement of the Spectral Index of the Diffuse Radio Background Between 90 and 190~MHz} 

\author[T. J. Mozdzen, J. D. Bowman, R. A. Monsalve, and A. E. E. Rogers]{T. J. Mozdzen$^{1}$\thanks{E-mail:
tmozdzen@asu.edu (TJM)}, J. D. Bowman$^{1}$, R. A. Monsalve$^{2,1}$, and A. E. E. Rogers$^{3}$\\
$^{1}$School of Earth and Space Exploration, Arizona State University (ASU), Tempe, Arizona, 85287, USA\\
$^{2}$Center for Astrophysics and Space Astronomy, University of Colorado (UCB), Boulder, Colorado, 80309, USA\\
$^{3}$MIT Haystack Observatory, Massachusetts Institute of Technology (MIT), Westford, Massachusetts, 01886, USA}

\begin{document}

\date{Accepted 2016 October 17. Received 2016 October 14; in original form 2016 August 24}

\pagerange{\pageref{firstpage}--\pageref{lastpage}} \pubyear{2016}

\maketitle

\label{firstpage}

\begin{abstract}
We report absolutely calibrated measurements of diffuse radio emission between 90 and 190 MHz from the Experiment to Detect the Global EoR Signature (EDGES). EDGES employs a wide beam zenith-pointing dipole antenna centred on a declination of $-26.7^\circ$. We measure the sky brightness temperature as a function of frequency averaged over the EDGES beam from 211 nights of data acquired from July 2015 to March 2016. We derive the spectral index, $\beta$, as a function of local sidereal time (LST) and find $-2.60~\textgreater~\beta~\textgreater~-2.62~\pm$0.02 between 0 and 12~h LST. When the Galactic Centre is in the sky, the spectral index flattens, reaching $\beta = -2.50~\pm$0.02 at 17.7~h. The EDGES instrument is shown to be very stable throughout the observations with night-to-night reproducibility of $\sigma_{\beta}$~\textless~0.003. Including systematic uncertainty, the overall uncertainty of $\beta$ is 0.02 across all LST bins. These results improve on the earlier findings of \cite{rog01} by reducing the spectral index uncertainty from 0.10 to 0.02 while considering more extensive sources of errors. We compare our measurements with spectral index simulations derived from the Global Sky Model (GSM) of \cite{deo01} and with fits between the \cite{guz01} 45~MHz and \cite{has02}~408~MHz maps. We find good agreement at the transit of the Galactic Centre. Away from transit, the GSM \textcolor{black}{tends to} over-predict \textcolor{black}{(GSM less negative)} by 0.05~\textless~$\Delta_{\beta}\textcolor{black}{ = \beta_{\text{GSM}}-\beta_{\text{EDGES}}}$~\textless~0.12, while the 45-408~MHz fits  \textcolor{black}{tend to} over-predict by $\Delta_{\beta}$~\textless~0.05.    
\end{abstract}


\section{INTRODUCTION}
The low-frequency radio sky spectrum between 50 and 200~MHz is a key area of interest because experiments seeking to detect redshifted 21~cm radiation from neutral hydrogen gas during the Epoch of Reionization (EoR) and earlier eras of First Light and X-ray heating must subtract astrophysical foregrounds in these frequencies to high-precision. At these frequencies, the sky is mostly dominated by Galactic synchrotron radiation but also contains contributions from supernova remnants and extragalactic radio sources.

There are two approaches to detecting the redshifted 21~cm signal.  The first approach is targeted by interferometeric arrays,  such as the Murchison Widefield Array (MWA;  \citealt{bow03, tin01}), the Precision Array to Probe the Epoch of Reionization (PAPER; \citealt{ali01}), the Hydrogen Epoch of Reionization Array (HERA; \citealt{pob01}), the Low-Frequency Array (LOFAR; \citealt{vha01}), and the Low-frequency Aperture Array component of the Square Kilometer Array (SKA; \citealt{mel01}).  These arrays aim to measure spatial fluctuations in the sky brightness temperature on arcminute and degree scales resulting from variations in the density, ionisation, and temperature of the intergalactic medium (IGM) above redshift $z>6$ (below 200~MHz).   

The second approach, on the other hand, exploits the bulk properties of the high-redshift IGM that yield a global (monopole) contribution of redshifted 21~cm signal to the all-sky radio background.  This approach is being pursued by the Experiment to Detect the Global EoR Signature (EDGES; \citealt{bow01,bow02}), Broadband Instrument for Global Hydrogen Reionisation Signal (BIGHORNS \citealt{sok01}), Sonda Cosmol\'{o}gica de las Islas para la Detecci\'{o}n de Hidr\'{o}geno Neutro (SCI-HI \citealt{voy01}), Large-aperture Experiment to detect the Dark Age  (LEDA \citealt{ber01}), Shaped Antenna measurement of background Radio Spectrum (SARAS \citealt{pat01, sin01}), and the Dark Ages Radio Explorer (DARE; \citealt{bur01}).  Both methods require careful and accurate subtraction of the sky foreground since the 21~cm signal is three to five orders of magnitude below the foreground.  

Although observations of the low-frequency radio sky were among the earliest in radio astronomy \citep{tur01, bri01}, detailed knowledge of the spectral properties of diffuse emission remains lacking.  The \citet{has01, has02} all-sky map at 408~MHz, which was compiled from data taken in the 1960s and 1970s, is still the cornerstone for foreground templates extrapolated across many frequencies in cosmic microwave background (CMB) and redshifted 21~cm analyses.  Improvements to the map have occurred throughout the years, including efforts by \citet{dav01, ben01, ben02, pla01}, and \citet{rem01} that focused on removing point sources and destriping the original data.  \citet{deo01} combined the Haslam 408~MHz map and several other surveys between 10~MHz and 94~GHz to build a Global Sky Model (GSM), which provides spectral properties of the diffuse emission of the entire sky. In this model at 150 MHz, the spectral index outside the Galactic plane is $\sim-2.6$ and in the Galactic plane (in a narrow band) increases to $\sim-2.5$ with peaks as high as $\sim-2.3$. Both the Haslam map and the GSM have become widely used and cited in both simulations and data reduction pipelines for redshifted 21~cm observations \citep{bow04, pat01, sub01, thy01, ber02}.

\citealt{guz01} created an all-sky temperature map at 45~MHz based upon the surveys of \citet{alv01} and \citet{mae01}. In addition, they produced an all-sky Galactic spectral index map based upon two frequency points by using their 45~MHz and the Haslam 408~MHz map after corrections for zero-level, extragalactic non-thermal emission, and CMB factors.

At higher frequencies, observations of Galactic radio emission between 3 and 90 GHz have been made by the Absolute Radiometer for Cosmology, Astrophysics, and Diffuse Emission I and II (ARCADE I and II) sky surveys \citep{fix01,kog01,sei01}. This survey constrained models of extragalactic emission and suggested that models of electric dipole emission from spinning dust particles were a possible explanation for excess emission seen at 1 cm wavelengths. They find that, at most, three parameters (reference brightness temperature,  spectral index, and curvature) are necessary to model their wide-band data combined with external radio measurements between 22 MHz and 1.4 GHz, and from WMAP at 23 GHz \citep{kog02}.

Recently, the new redshifted 21~cm arrays have begun yielding the first large, multi-spectral maps sensitive to diffuse structures in the low-frequency sky, including recent surveys by MWA GLEAM \citep{way01} and LOFAR \citep{hea01}. GLEAM will scan the entire sky south of $\delta \sim+25^{\circ}$ between 72 and 231 MHz, and the LOFAR survey will scan 100 sq degrees in the northern sky centred on $(15^h, 69^{\circ})_{\text{J2000}}$ between 30 and 160 MHz. The LOFAR survey's primary purpose is to enable automated processing by providing an \textit{a priori} sky model, which also serves as an aid to foreground removal for EoR detection. Similarly, GLEAM's survey will serve a myriad of uses, one of which again will be foreground removal in EoR searches. 

The EDGES instrument is able to provide a unique measurement of the absolute sky brightness temperature averaged over large spatial scales on the sky due to its wide beam.  \citet{rog01} found that the spectral index,  $\beta_{100-200}$,  of diffuse emission,  defined as $T_\text{sky}\propto \nu^{\beta}$,  was $-2.5\pm0.1$ at high-Galactic latitudes in the frequency range 100-200~MHz. Those measurements were taken using three days of data; two of which used a N-S orientation of the antenna's excitation axis and one with an E-W orientation. By combining their results with the Haslam sky map at 408~MHz, they were able to show that  $\beta_{150-408} = -2.52\pm0.04$ at high Galactic latitudes.

In this paper we present new observations taken over a span of 240 nights from July 2015 through March 2016 with the latest version of the EDGES instrument that deploys an improved antenna with better chromatic performance and a new high-precision receiver calibration approach.  These advancements enable us to improve on our earlier measurements of the spectral index of the diffuse low-frequency radio emission and extend our coverage to all sidereal times at a constant declination of $-26.7^\circ$.  

The paper is organized in the following manner.  In Section 2 we describe the instrument and calibration details. In Section 3 we present details of the data collected and chromatic beam corrections. Section 4 presents and discusses the spectral index results and comparisons to values predicted by relevant sky models.
\section{EDGES Instrument}
\label{sec:instrument}

The EDGES experiment is deployed at the Murchison Radio-astronomy Observatory (MRO) in Western Australia (-26.7$^\circ$,~+116.6$^\circ$).  The experiment consists of two scaled-replica instruments spanning neighboring frequency bands of 50-100~MHz and 90-190~MHz.   Each instrument consists of a single dipole-based antenna and receiver.  Extensive modeling and simulation is performed on all instrument components, along with calibration measurements conducted both in the laboratory prior to instrument deployment and on site following deployment, in order to meet the low systematic instrumental error requirements for redshifted 21~cm observations \citep{rog02}.   For the results reported here, we used the high-band instrument, hence we focus our discussion on that instrument.  In this section, we review the instrument and the primary calibration steps.  

\subsection{Antenna and Receiver} 

\begin{figure}
  \includegraphics{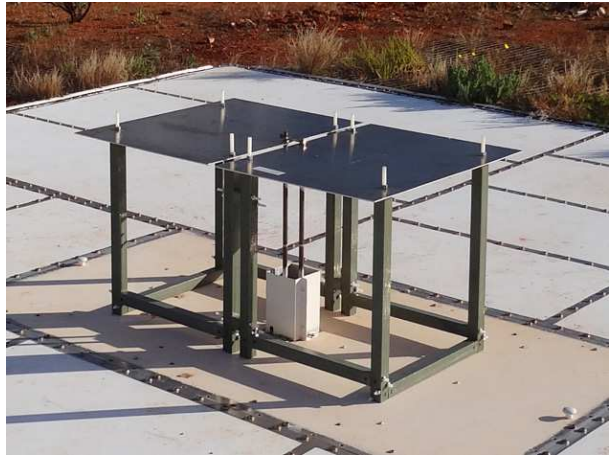}
 \caption{(left) Photograph of the high-band blade antenna with fiberglass support tubes and a tuning capacitor on the top of the panels between the balun tubes to improve impedance matching. Surrounding the balun tubes at the base, a short rectangular enclosure shields against vertical currents in the tubes.}
  \label{fig:antenna_images}
\end{figure}

The high-band instrument employs a single broadband ``blade'' dipole antenna sensitive to wavelengths in the range $3.3\geq\lambda\geq1.6$~m ($90\le\nu\le 190$~MHz).  The antenna is ground-based and zenith pointing.  The antenna consists of two rectangular planar panels that are each 63~$\times$~48~cm$^2$ and are placed 52~cm above a ground plane.  The metallic ground plane is formed using a 5.35~$\times$~5.35~m$^2$  aluminum plate underneath the antenna, with four wire mesh extensions (each 2~$\times$ 5~m$^2$) that combine to form a ``plus'' shape. The antenna uses a Roberts Balun \citep{rob01} with a small shield at the bottom (Fig.~\ref{fig:antenna_images}).  The blade antenna replaces the older ``fourpoint'' antenna design used in prior EDGES studies and yields better chromatic performance by introducing less spectral structure into the measurement \citep{moz01}.    The full width at half maximum beamwidth of the antenna at 150~MHz is 72$^\circ$ parallel to the axis of excitation and 110$^\circ$ perpendicular to the axis. The features of the antenna are summarized in Table~\ref{tab:antennas}.

\begin{figure}
  \includegraphics{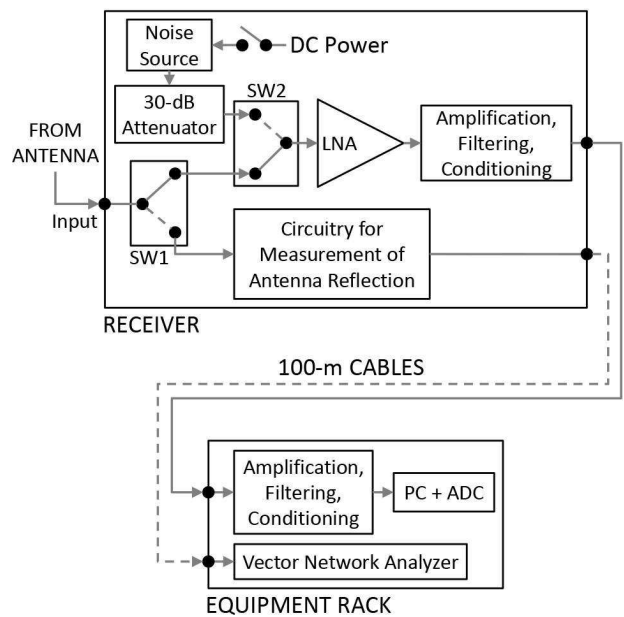}
 \caption{Block diagram of the EDGES instrument}
  \label{fig:block_diagram}
\end{figure}

The antenna is connected to a receiver placed underneath the ground plane in a temperature-regulated enclosure kept at $25^{\circ}$C. The receiver amplifies and conditions the signal before passing it through a 100~m cable to a backend unit that further amplifies it and sends it to a PC-based 14-bit 400-MS/s analog to digital converter (Fig. \ref{fig:block_diagram}). Blocks of 65536 voltage samples are then Fourier transformed to obtain spectra of 32768 points below 200~MHz with 6.1~kHz resolution.  

\subsection{Calibration and Corrections} 

The receiver utilizes laboratory calibration prior to deployment, augmented with a three-position hot/cold/antenna calibration switching scheme for stability in the frontend during operation, to achieve an absolute accuracy of $\le0.1$\% in measured antenna temperature \citep{mon01}. The calibration quantities are measured as a function of both temperature and frequency. The impedance match of the antenna can be measured \textit{in situ} by switching a Vector Network Analyzer (VNA) into the electrical path.

To propagate the calibration to the sky temperature measurement, we must apply several corrections for losses that occur in the antenna before the receiver.  The losses include ground plane loss, balun loss, and antenna panel loss. We also make an estimate and correct for the actual temperature of the low-noise amplifier (LNA) which may deviate from the nominal 25$^\circ$C due to temperature gradients.

Ground plane loss originates from our finite ground plane, allowing a small amount of signal ($\le1\%$) from below the horizon to be received from the antenna. These losses are modeled through CST (Computer Simulation Technology) Microwave Studio and through FEKO (FEldberechnung f\"{u}r K\"{o}per mit beliebiger Oberfl\"{a}che) E\&M simulations. The balun loss is due to minor resistive impedance in the brass and copper-plated balun hardware and also to the resistivity and dielectric properties of the connector to the receiver. This effect is estimated from analytical models.  Finally, losses in the antenna panels are due to finite conductivity and estimated with FEKO. Adjustments are made to the measured spectrum to counteract these effects in processing.

\begin{table}
   \caption{Blade Antenna Features (refer to Fig. \ref{fig:antenna_images})}
   \label{tab:antennas}
   \begin{tabular} {l  l}
   	\hline
  	Parameter & Value\\
   	\hline
  	3 dB Beamwidth $\phi$=~0$^{\circ}$ at 150 MHz &72$^{\circ}$\\
  	3 dB Beamwidth $\phi$=90$^{\circ}$ at 150 MHz &110$^{\circ}$\\
 	Height above ground plane &52 cm\\
	Panel Width &63 cm\\
	Panel Length &48 cm\\
	Solid ground plane&5.35~$\times$~5.35 m$^2$\\
	Mesh ground plane extensions&2~$\times$~5 m$^2$\\
     \hline
	&\\
	&\\
	&\\
	&\\
	\end{tabular}
\end{table}
\section{Data and Processing}
We measured the sky temperature using EDGES over a period of 240 days starting with day 207 (July 26) of 2015. We determined the spectral index by fitting the calibrated data to the two parameter equation
\begin{equation} 
\label{eq:Power_Law_1}
T_\text{sky} = T_{150}\left(\frac{\nu}{\nu_0}\right)^{+\beta} + T_\text{CMB},
\end{equation} 
where the two fitting parameters, $\beta$ and $T_{150}$, are the spectral index and the brightness temperature at 150~MHz less $T_\text{CMB}$, respectively. The CMB temperature, $T_\text{CMB}$, is 2.725~K, and $\nu_0$ is 150~MHz. We find that this equation yields lower RMS fitting errors than the log form of the equation: 
\begin{equation} 
\label{eq:Power_Law_2}
\text{ln}(T_\text{sky}-T_\text{CMB}) =\text{ln}(T_{150}) + \beta \text{ln}\left(\frac{\nu}{\nu_0}\right) .
\end{equation} 

Data collection was occasionally interrupted due to a variety of factors such as power outages and inclement weather (Table~\ref{tab:missing_data}). We exclude daytime data since effects from the sun and the ionosphere \citep{rog03} can interfere with the observation of the astronomical sky foreground. Sporadic data points affected by radio-frequency interference (RFI) are flagged and excised. The data are averaged and binned into 72~timeslots of 20~minutes in LST and into 250 frequency bins of 400~kHz from 90~to 190~MHz.

For an antenna on the surface of the Earth pointed toward zenith, the pointing declination of its primary axis corresponds to the latitude of the antenna's deployment site, and the pointing right ascension corresponds to the LST at the site. As the earth rotates, the antenna pointing changes direction, altering the mapping of its beam pattern onto sky coordinates. Because the antenna's beamwidth is wide by design, we sample the average sky temperature weighted by the beam (see Figs. \ref{fig:beam_width} - \ref{fig:galactic_beams}).

At low frequencies, the Earth's ionosphere can distort the appearance of the astronomical sky and, in principle, provide an additional chromatic effect.  \cite{ved01} used the LOFAR low frequency antenna beam to evaluate the chromatic effects of the ionosphere, in particular due to refraction. To minimize these effects, we restrict our data analysis to times when ionospheric perturbations are minimal by excluding data when the sun's elevation is -10$^{\circ}$ or higher.
\begin{table}
  \center
   \caption{Days of omitted data between day 207 of 2015 and day 82 of 2016. Out of 240 observing days, 29 days did not capture usable nighttime data.}
   \label{tab:missing_data}
   \begin{tabular} {l  l  c}
   	\hline
  	Day Numbers&Year&Span\\
   	\hline
	208 to 209 & 2015&2\\
  	212 to 214 &&3\\
	244 & &1\\
	246 to 249 && 4\\
	263 to 264 && 2\\
	290 & & 1 \\
	342 && 1\\
	303~to 309 && 7\\
	 ~20~to  ~26&2016&7\\
      ~54&2016&1\\
	\hline
	Total Days & & 29\\
     \hline
	\end{tabular}
\end{table}
\subsection{Adjustment for Beam Chromaticity} 
In the context of measurements of the sky spectrum, an ideal antenna beam would be independent of frequency.  However, as shown in \citet{ber01,moz01}, realistic antennas suffer chromatic (frequency-dependent) effects that couple angular structure in the sky to spectral structure in the measurement.  Thus, before we can analyze the spectral properties of the radio sky to high-precision, we must correct for the chromatic effects of the EDGES antenna.

\begin{figure}
  \includegraphics{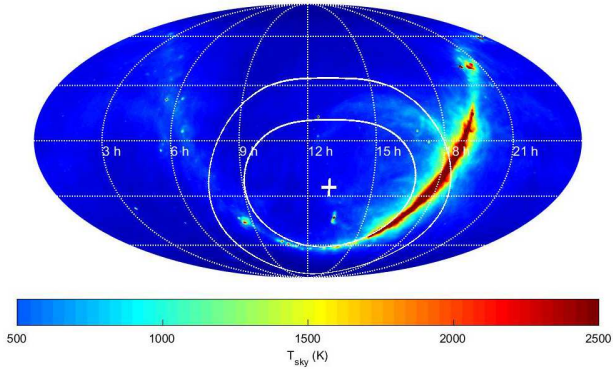}
 \caption{The EDGES beam at -3 and -10 dB projected onto the Haslam Sky map scaled to 150~MHz in Celestial coordinates when located at -26.7$^\circ$ latitude with the zenith at 13~h LST. The beam is purposefully wide to capture the sky averaged global EoR signal.}
  \label{fig:beam_width}
\end{figure}

\begin{figure}
  \includegraphics{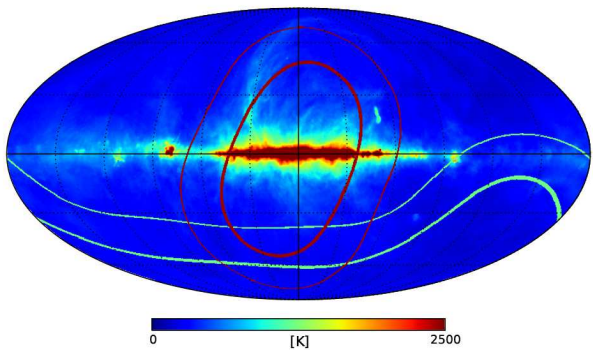}
 \caption{The EDGES beam at -3 and -10 dB projected onto the Haslam Sky map scaled to 150~MHz in galactic coordinates when located at -26.7$^\circ$ latitude shown for LST values of 18~h and 2.5~h (dark trace and light blue trace respectively). The beam at 18~h LST captures much of the central region the Galactic plane, which is high in spatial structure, while at 2.5~h it captures mostly structureless sky, minimizing chromatic beam effects.}
  \label{fig:galactic_beams}
\end{figure}

Fig.~\ref{fig:beam_width} shows a snapshot of the beam's directivity at 150~MHz projected upon the Haslam sky map when located at -26.7 latitude and pointed towards the zenith at LST~=~13~h. When the directivity of the beam changes at other frequencies, affected locations in the sky will contribute stronger or weaker to the antenna temperature. These chromatic beam effects are significant especially with the Galactic Centre overhead where the RMS fitting errors to a two parameter equation can exceed 9~K for an uncorrected spectrum. \cite{rog01} chose to report the spectral index using a region in the sky which would minimize chromatic beam effects and chose LST~=~2.5~h. Fig.~\ref{fig:galactic_beams} shows that this is true for the present blade beam, but even in this region, beam effects must be taken into account. Compared to other antenna choices, we have significantly reduced this effect by using the blade design, which has the benefits of minimal chromaticity and is conducive to precision beam modeling through simulations \citep{moz01}.  

\begin{figure}
\centering
  \includegraphics{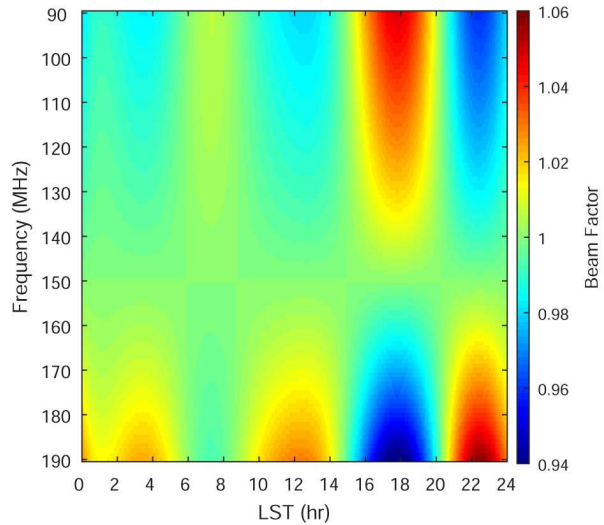}
  \caption{Simulated beam chromatic correction factor for the EDGES blade antenna. For a given LST, the beam factor is created by dividing the simulated convolution of the chromatic beam with the sky at 150 MHz, by the simulated convolution of both the beam and the sky at 150 MHz. This ratio gives the amount of over or under antenna temperature as compared to the antenna temperature if the beam were not chromatic and had a directivity pattern of the beam at 150~MHz. The magnitude of the beam factor is greatest when there is angular structure in the sky as is the case near LST 18 h when the Galaxy is overhead.}
  \label{fig:beam_factor}
\end{figure}

To further reduce the chromatic beam effect in measured spectra, we use simulated beam maps, along with a sky model, to generate a chromaticity correction factor for the beam, which can reduce the RMS fitting errors to under 5~K.  We model the antenna from 90~to~190~MHz in steps of 1~MHz with numerical time-domain electromagnetic simulations using CST Microwave Studio.  We then use these beam models, along with the Haslam 408 MHz map, to compute a correction factor as a function of frequency for each LST in our observations.  To scale the Haslam map from 408~MHz to our frequency range, we first subtract the $T_{\text{CMB}}$ temperature from the map, scale its brightness temperature to 150 MHz using a constant spectral index of $\beta=-2.5$ and then add $T_{\text{CMB}}$ back in.  We convolve the scaled Haslam map with the beam directivity at each frequency modeled.  The beam factor ratio is formed by dividing the modeled antenna temperature at each frequency by the antenna temperature at 150~MHz.  We implicitly assume that the spectral index does not vary significantly from its value at 150 MHz in our frequency range of 90~ to 190~MHz. The beam correction factor is given by:
\begin{equation} 
\label{eq:beam_correction}
B_{\text{factor}}(\nu) = \frac{\int_{\Omega}T_\text{sky}(\nu_{150},\Omega) B(\nu,\Omega) \mathrm{d}\Omega}{\int_{\Omega}T_\text{sky}(\nu_{150},\Omega) B(\nu_{150},\Omega) \mathrm{d}\Omega } ,
\end{equation}
where
\begin{equation} 
\label{eq:sky_correction}
T_\text{sky}(\nu_{150},\Omega) = \left [ T_\text{Haslam}(\Omega) - T_\text{CMB}\right ] \left ( \frac{150}{408} \right )^{-2.5}  + T_\text{CMB} ,
\end{equation}
$B(\nu,\Omega)$ is the beam directivity, $T_\text{Haslam}(\Omega)$ is the Haslam sky map, $T_\text{CMB}$ is 2.725~K, and $\nu$ is the frequency.
This ratio is the chromatic beam factor that we apply to the measured spectrum as a divisor to counteract the under or over response to the sky. We repeat this for all 72~time steps in LST to form a two-dimensional beam factor array (see Fig.~\ref{fig:beam_factor}).

\section{RESULTS}
\label{sec:results}
\subsection{Spectral index} 

From day number 207 in 2015 to day 82 in 2016, data were fit separately per day for each LST interval. The parameters $T_{150}$, $\beta$, and the RMS error to the fit were extracted using equation~(\ref{eq:Power_Law_1}).  The spectra were fit for the two cases of using or not using beam chromaticity correction.

\begin{figure}
\centering
   \includegraphics{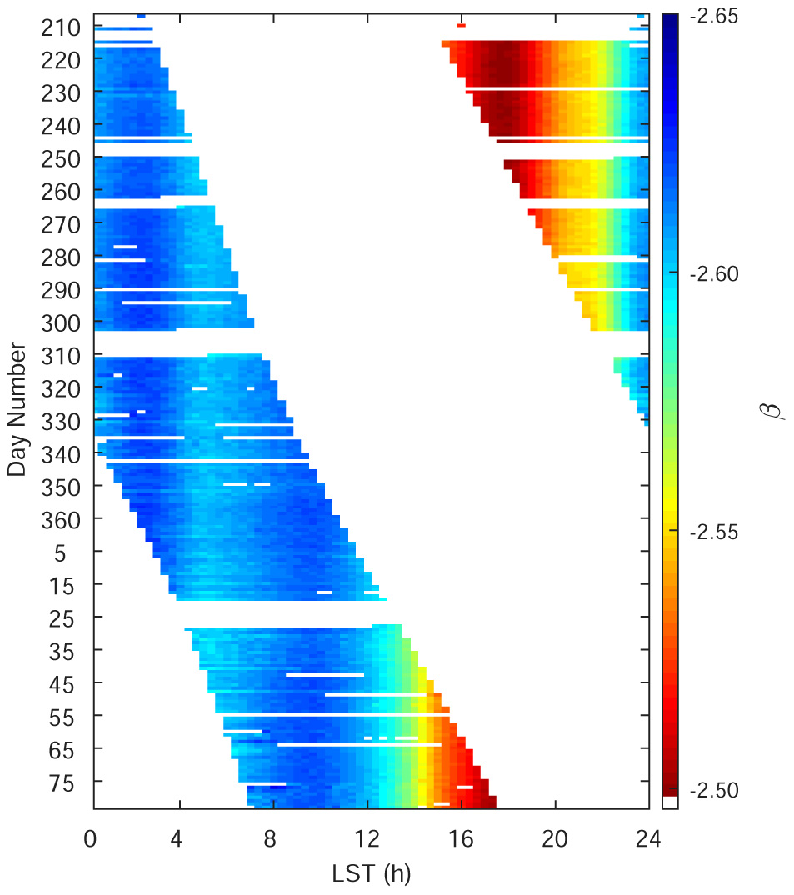}
   \includegraphics{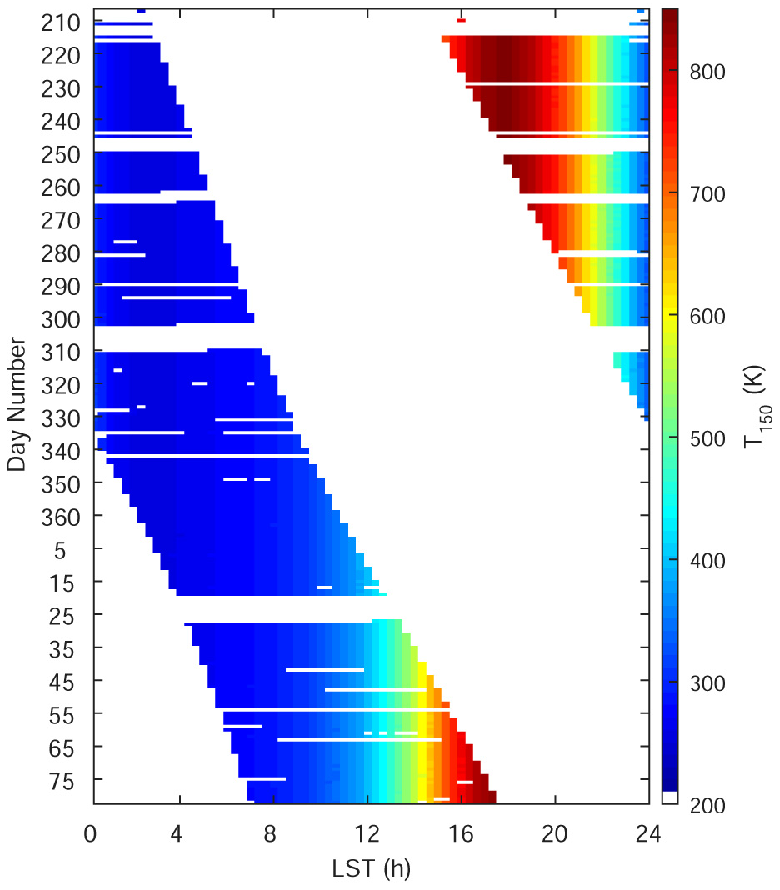}
   \caption{Waterfall graphs of (top) the spectral index $\beta$ and (bottom) the sky temperature at $\nu_0$=150~MHz, $T_{150}$. RFI and other spurious signals were purged from the spectra before performing a fit to equation \ref{eq:Power_Law_1}. The data were binned into 400~kHz wide bins from 90~to~190 MHz and the beam adjustment factor was applied. Both the $\beta$ and the $T_{150}$ graphs show excellent stability from day to day as the  data collection ran for 240 days from day 207 2015 to day 82 2016.}
  \label{fig:Waterfalls_two_params}
\end{figure}

\begin{figure}
\centering
   \includegraphics{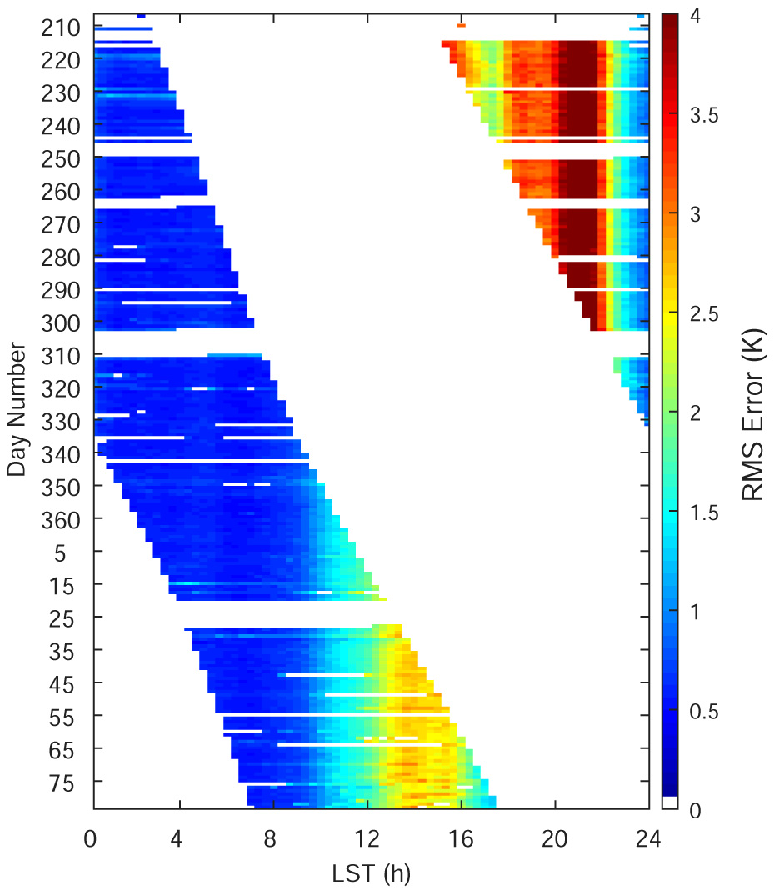}
   \caption{Waterfall graph of the RMS fitting error to equation~(\ref{eq:Power_Law_1}). RFI and other spurious signals were purged from the spectra before performing the fit. The data were binned into 400~kHz wide bins from 90 to 190 MHz and the beam adjustment factor was applied. The RMS error also shows excellent stability from day to day as the data collection ran for 240 days from day 207 2015 to day 82 2016.}
  \label{fig:Waterfalls_two_params_RMS}
\end{figure}

\begin{figure}
 \centering
 \includegraphics{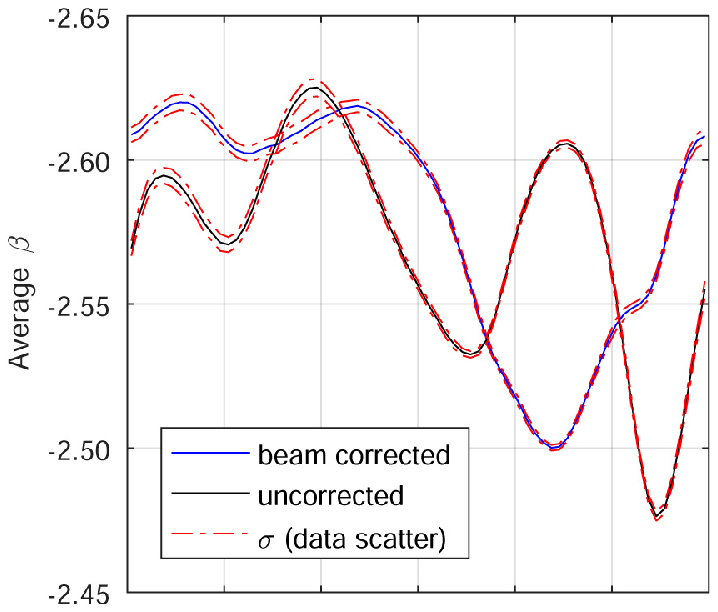}
 \includegraphics{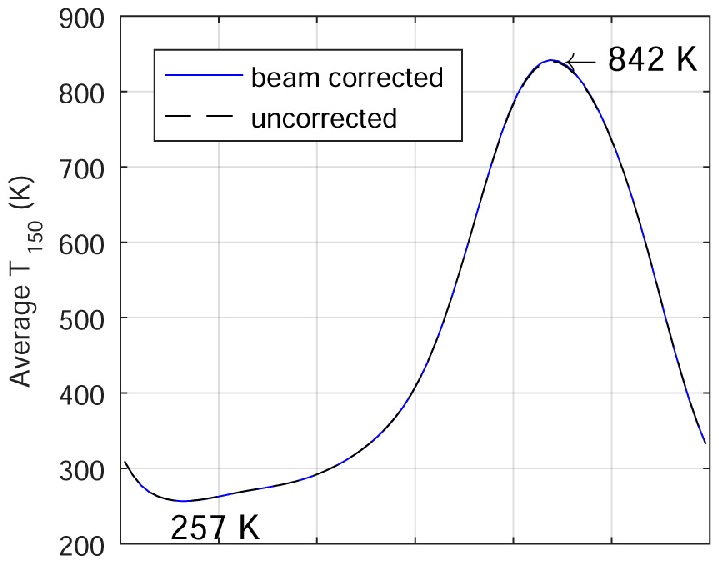}
 \includegraphics{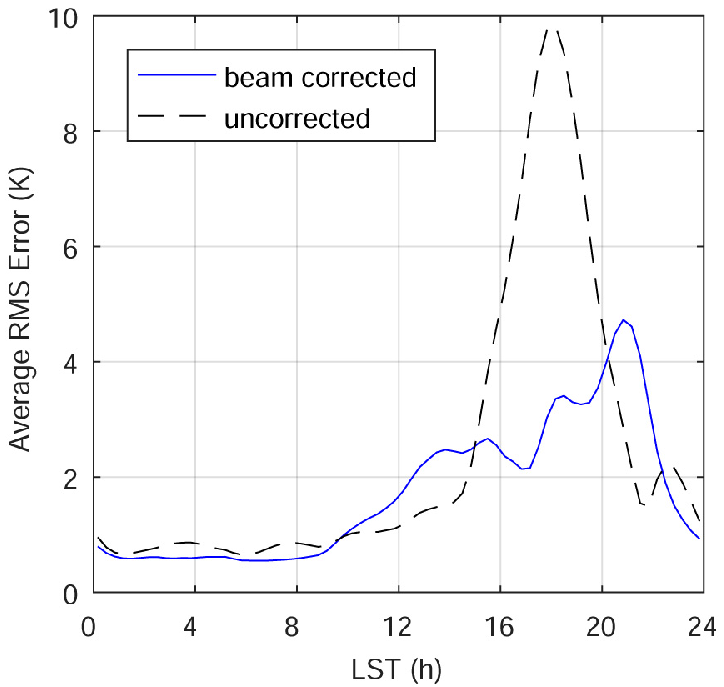}
 \caption{Averages of $\beta$, $T_{150}$, and RMS error for each LST value taken over days for which nighttime data was available. Averages are shown for the processing done with and without beam correction. Beam correction reveals the dip in $\beta$ near the Galactic Centre, where the variations of $\beta$ without beam correction are dominated by the chromatic beam effects. The small data scatter in $\beta$ further confirms the uniformity from day to day seen in the waterfall graphs. $T_{150}$ min and max values, 257~K and 842~K, are annoted on the chart.}
 \label{fig:Grouped_Averages_two_params}
\end{figure}

The values of $T_{150}$, $\beta$, and the RMS error show excellent stability over time as can be seen in the waterfall plots of Figs.~\ref{fig:Waterfalls_two_params}-\ref{fig:Waterfalls_two_params_RMS} and confirmed by the low standard deviation in data scatter as seen in the $\beta$ and $T_{150}$ averages of  Fig.~\ref{fig:Grouped_Averages_two_params}, where $\sigma_{\beta}$~\textless~0.003 and $\sigma_{T_{150}}$~\textless~5~K. $T_{150}$ is also insensitive to beam chromaticity as the average beam corrected and uncorrected values differ by no more than 0.2\% due to the low chromaticity of the blade beam. The daytime data (sun above $-10^\circ$ elevation) are not shown, yielding the blank band in the plot.

The importance of including the beam correction factor is most apparent in the average $\beta$ plot of Fig.~\ref{fig:Grouped_Averages_two_params}. With the correction applied, we find $-2.60~\textgreater~\beta~\textgreater~-2.62$ for LST values between 0~and 12~h. Between 12 and 24~h it features a flattening that peaks at $-2.50$ at 17.7~h. It is clear in Fig.~\ref{fig:Grouped_Averages_two_params} that beam correction has a significant impact on the recovered spectral index.  Without correction, the beam effects mask the flattening that is expected near the Galactic Centre \citep{deo01, guz01}. 

We also investigated fitting the spectra to a three parameter model, adding $\gamma$ as a third term 
\begin{equation} 
\label{eq:Power_Law_2}
\text{ln}(T_\text{sky}-T_\text{CMB}) =\text{ln}(T_{150}) + \beta \text{ln}\left(\frac{\nu}{\nu_0}\right)+ \gamma \left[\text{ln}\left(\frac{\nu}{\nu_0}\right)\right]^2 .
\end{equation} 
Although the RMS fitting errors were smaller and $|\gamma|$ was typically relatively small, \textless~0.1, we conservatively assume that the uncertainty on gamma due to systematic effects dominates fitted values.  Hence, we save further three parameter analysis to future work.

\begin{figure}
 \centering
 \includegraphics{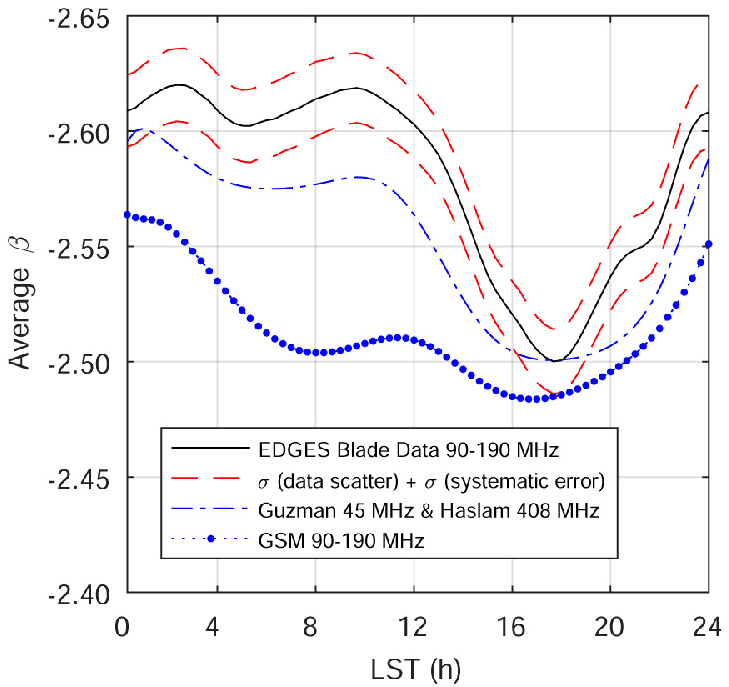}
 \caption{Spectral index derived from EDGES measurements including the total one-sigma uncertainty, which consists of 0.012 for systematic uncertainty and the LST dependent data scatter. Also plotted are the simulated spectral index results when using the GSM between 90 and 190~MHz and by the combination of the Guzm\'an (45 MHz) and Haslam (408 MHz) maps. The two simulated spectral index curves show similar structure in $\beta$ as compared to the measurement derived curve, but the Guzm\'an curve matches the measured data more closely.}
 \label{fig:GSM_vs_Guz-Has}
\end{figure}

\subsection{Systematic error estimation} 
We evaluate and correct for sources of systematic errors to the calibration process: LNA temperature gradients, ground loss, antenna loss, and beam correction uncertainty. We maintain the receiver at $25^{\circ}$C with forced airflow and active cooling/heating, but the temperature of the LNA's electronic components may differ slightly from the reading at the main temperature sensor. A second sensor is used to capture potential gradients, which remain below $1^{\circ}$C during operation. This temperature difference has a small effect on the calibration values that are used to adjust the antenna spectrum. Simulations showed the impact on $\beta$ to be no more than 0.005 if we did not include the adjustment.

In the field, we use a $5.35~\times~5.35$~m$^2$ solid metallic ground plane made from a central square of solid aluminum augmented by $2~\times~5$~m$^2$ sheets of wire mesh. The imperfect ground of the earth serves as the ground plane where there is no metal. Given the limited size of the metal ground plane, a small fraction of the antenna beam will pickup the temperature of the ground. The effect of the finite ground plane on ground pickup/loss was simulated in FEKO and CST. In FEKO we used a Greens function to implement a finite ground plane modeled over the soil.  We used values for the dielectric constant of 3.5 and conductivity of 0.02~S/m because they are based upon the soil measurements made at the MRO for dry conditions reported by \cite{sut01}. In CST we modeled the finite ground plane as a finite sheet in free space. Both simulations resulted in losses $\textless$~1.0\% but with opposite frequency dependence trends. CST results varied between 0.4 and 1.0\%, with the larger loss at lower frequencies, while FEKO varied between 0.2 and 1.0\% with larger loss at high frequencies. The effects are small in both cases and, considering the simulation uncertainties, it is not clear which model is more appropriate. Hence, we took the approximate average of the two simulations as a constant 0.5\% loss, indepdendent of frequency. Correcting for 0.5\% frequency-independent ground loss caused $\beta$ to become more positive by up to 0.003 for 0~h~$\textless$~LST~$\textless$~12~h and become more positive by 0.003-0.006 for 12~h~$\textless$~LST~$\textless$~24~h as compared to no ground loss correction.

The antenna losses arise from the balun resistance, the connection to the receiver, and the resistance of the antenna panels. The balun of the EDGES system includes two brass tubes (outer diameter 0.5") with a copper-plated brass rod running up one of the tubes and a teflon-dielectric connector to the receiver. When we account for the impedance in the balun structure and its connection to the receiver in the calibration process (parameters obtained via transmission line models and lab measurements), we see an effect on $\beta$ no greater than 0.015. There is a very minor effect on $\beta$ from the panel resistance of 0.001.

When we use CST and FEKO to compute the beam correction factor, the difference in the spectral index between methods is 0.016. This considers different choices of including or subtracting the CMB as well as different scalings of the Haslam 408~MHz map to 150~MHz assuming realistic spectral indices. To assess the impact of spatial variations in the spectral index on our beam correction factor, we used the frequency dependent GSM instead of the spatially flat scaled Haslam map. The results show that the GSM-derived beam correction factor differs from our default method by 0.3\% at most. The effect upon the spectral index causes a difference of $\pm$0.003 for all LST values except from 20 to 23 h where it rises to +0.005. Hence, the main component of chromaticity originates from the beam and not spectral index spatial variations.

We now combine the sources of systematic uncertainty. If we assume our adjustments have a standard error of half of the magnitude of the effect (with vs. without or method 1 vs method 2), and combine the sources of system error in quadrature: temperature gradients 0.005, finite ground plane effects 0.006; antenna balun 0.015; finite panel resistance 0.001; and beam correction factor 0.016, we arrive at a systematic uncertainty on $\beta$ of 0.023/2~=~0.012. Adding the maximum data scatter value of 0.003 directly to the systematic uncertainty, as opposed to in quadrature, the total maximum uncertainty in  $\beta$ becomes 0.015. Based on this value we conservatively report a total one-sigma uncertainty of $\sigma_{\beta_\text{Total}}~=~0.02$. This uncertainty improves upon the uncertainties reported in \cite{rog01} of $\sigma_{\beta_\text{100-200}}~=~0.1$ considering only EDGES data and $\sigma_{\beta_\text{150-408}}~=~0.04$ when including Haslam skymap data. Additionally, we measure data over a significantly longer time period and include the LST dependence of the spectral index. The final measurement with the full uncertainty range is shown in Fig.~\ref{fig:GSM_vs_Guz-Has}.

\subsection{Simulated spectral index from sky maps} 
We compare our measured results for the spectral index to simulated results by using publicly available sky maps: the de Oliveira-Costa GSM; the Haslam 408~MHz map; and the Guzm\'an 45~MHz map. We simulate the antenna temperature that EDGES is expected to observe using the GSM from 90 to 190~MHz in steps of 1~MHz as a function of LST, by convolving the sky model (less $T_{\text{CMB}}$) with the simulated EDGES blade beam (fixed at 150~MHz to compare to the beam-corrected measurements). We then fit the simulated antenna temperature to equation~(\ref{eq:Power_Law_1}) to obtain the spectral index vs. LST.  Because the Haslam and Guzm\'an sky maps are only available at 408~MHz and 45~MHz, respectively, we calculate the spectral index with the closed form expression
\begin{equation} 
	\label{eq:Two_frequency_spectral_index}
	\beta =\text{ln}\left(\frac{T^{'}_{\text{ant}}(\nu_{45})}{T^{'}_{\text{ant}}(\nu_{408})}\right) \bigg/ \text{ln}\left(\frac{\nu_{45}}{\nu_{408}}\right),
\end{equation} 
where $T^{'}_\text{ant}(\nu)$ is
\begin{equation} 
\label{eq:T_antenna}
T^{'}_\text{ant}(\nu) = \int_{\Omega}T^{'}_\text{sky}(\nu,\Omega) B(\nu_{150},\Omega) \mathrm{d}\Omega ,
\end{equation}
$B(\nu_{150}, \Omega)$ is the achromatic beam for a given pointing and orientation, $\nu$ is frequency,  and $T^{'}_\text{sky}(\nu,\Omega)$ is the sky model in use.  $B(\nu_{150}, \Omega)$ is normalized to a unit integral at each frequency. In this manner, we generate spectral index values across 24 hours of LST. The results of these two simulations along with our result from measured data are shown in Fig.~\ref{fig:GSM_vs_Guz-Has}.  

We find that the GSM-derived spectral indices  tend to over-predict \textcolor{black}{(GSM less negative)} the spectral index of the observations at high-Galactic latitudes (low LST). The GSM-derived spectral indices differ from our measurements by 0.05~\textless~$\Delta_{\beta}\textcolor{black}{=\beta_{\text{GSM}}-\beta_{\text{EDGES}}}$~\textless~0.12 (3~to~7.5~sigma) away from the Galactic Centre and $\Delta_{\beta}~=~0.01$ (1~sigma) near the centre.  The Guzm\'an-Haslam (GH) derived spectral indices, on the other hand, are a closer match.  The GH indices only differ from our measurements by  0.01~\textless~$\Delta_{\beta}$~\textless~0.05 (1~to~3~sigma) away from the Galactic Centre and agree well at 18 h LST.  One possible cause for the better agreement to the data by the GH model compared to the GSM is the higher sky temperatures reported in the Guzm\'an map as compared to the GSM at 45 MHz. Away from the Galactic Centre, the Guzm\'an map is 1100 K higher than the GSM in much of the sky. If the GSM had higher temperatures away from the Galactic Centre, the simulated spectral index using the GSM maps would decrease for the lower values of LST, giving better agreement with our measurements.   Based on these results, we caution that the GSM likely under-predicts the sky temperature at high-Galactic latitudes below 200 MHz.

If we make a direct comparison against \cite{rog01} where their EDGES-only spectral index result was $-2.5~\pm~0.1$ at LST~=~2.5~h, we report $-2.62~\pm~0.02$, which is consistent with the 2008 results. To compare against their $\beta_{150-408}$ value, we must use our $T_{150}$ measurement values at LST~=~2.5~h coupled with the $T_{408}$ values as computed using the Haslam sky map with our antenna beam. In this case, we compute $\beta_{150-408}=-2.63~\pm~0.01$ as compared to their result of $\beta_{150-408}=-2.52~\pm~0.04$. The disagreement in these two values arise from differences in $T_{408}$ and $T_{150}$ values. $T_{408}$ values differ by 0.7 K, which is most likely due to different antenna beam models in use (due to different antenna designs) and results in 33~\% of the disagreement.  $T_{150}$ values differ by twice the 2008 stated uncertainty in $T_{150}$ ($\sigma_{T_{150}}$~=~10~K), which we believe is due to accuracy of the \cite{rog01} measurement and accounts for the remainder of the difference.

\section{CONCLUSION}
\label{sec:conclusion}
We measured the sky brightness temperature as a function of frequency and derived the spectral index $\beta$ as a function of sidereal time by fitting to a two parameter model over a span of 240 days using 211 days of nighttime data acquired from July 2015 to March 2016. Instrument calibration, including corrections for temperature gradients, ground loss, antenna losses, and beam chromaticity, has been applied to deliver instrument stability of over several months as demonstrated by spectral index standard deviation values $\sigma_{\beta}~\textless$~0.003.

We have presented results of extensive measurements of the diffuse radio sky between the frequencies of 90-190~MHz. We find that the spectral index $\beta$ is in the range $-2.60~\textgreater~\beta~\textgreater~-2.62~\pm$~0.02 in the lower LST values, but increases to $-2.50$ with the Galaxy overhead. The GSM tends to over-predict the strength of the variation in the spectral index in the range 0.05~\textless~$\Delta_{\beta}$~\textless~0.12 for low LST. However, comparison with the spectral index as computed using the Guzm\'an sky map at 45 MHz and the Haslam Sky map at 408 MHz is a closer match and differs by  0.01~\textless~$\Delta_{\beta}$~\textless~0.05 away from the centre. At the Galactic Centre both models agree with our measurements of the spectral index to within one sigma.

Future work is planned to investigate the criteria needed to fit to a three-parameter spectral equation. Updates to sky models will be used to simulate spectral index across LST using the EDGES blade antenna. We also plan to measure the spectral index in the $50 - 100$~MHz range with the low-band EDGES blade antenna.

\section*{Acknowledgments}
The authors would like to thank Adam Beardsley for useful inputs and discussions. This work was supported by the NSF through research awards for the Experiment to Detect the Global EoR Signature (AST-0905990 and AST-1207761) and by NASA through Cooperative Agreements for the Lunar University Network for Astrophysics (NNA09DB30A) and the Nancy Grace Roman Technology Fellowship (NNX12AI17G). Raul Monsalve acknowledges support from the NASA Ames Research Center (NNX16AF59G). EDGES is located at the Murchison Radio-astronomy Observatory. We acknowledge the Wajarri Yamatji people as the traditional owners of the Observatory site. We thank CSIRO for providing site infrastructure and support.

\bsp

\label{lastpage}
\end{document}